\documentclass[11pt,twoside]{article}

\usepackage{asp2006}
\usepackage{epsf}
\usepackage{graphicx}
\usepackage{lscape}
\usepackage{amssymb}

\begin{document}
\title{The Magnetic Sensitivity of the Second Solar Spectrum}   %%% Fill in title
\author{J. Trujillo Bueno}   %%% Fill in author names
\affil{Instituto de Astrof\'{\i}sica de Canarias, 38205, La Laguna, Tenerife, Spain}    %%% Fill in author affiliations

\begin{abstract} %%% Abstract to run on from here.
This paper reviews some of the developments that over the last 10 years have allowed us to go from deciphering the physical origin of several of the enigmatic features of the second solar spectrum to discovering unknown aspects of the Sun's hidden magnetism via sophisticated radiative transfer modeling. The second solar spectrum is the observational signature of radiatively induced quantum coherences in the atoms and molecules of the solar atmosphere. 
Magnetic fields produce partial decoherence via the Hanle effect, giving rise to fascinating observable effects in the emergent spectral line polarization. Interestingly, these effects allow us to ``see" magnetic fields to which the Zeeman effect is blind within the limitations of the available instrumentation. In the coming years, the physical interpretation of observations of the spectral line polarization resulting from the joint action of the Hanle and Zeeman effects might lead to a new revolution in our empirical understanding of solar magnetic fields.

\end{abstract}

\section{Historical Introduction}  

\begin{quote}
{\small ``{\em Recent accurate measurements of the wavelength dependence of linear polarization of the solar limb radiation has resulted in the discovery of what may be called the second solar spectrum, with dozens of polarization ($Q/I$) features seen both `in absorption' and `in emission', i.e., with correspondingly smaller and larger polarization than in the adjacent continuum.}"  (Ivanov 1991)}.
\end{quote}

Before 1980 ``non-magnetic" scattering line polarization had been detected only in a few spectral lines. Except perhaps for the Ca {\sc i} line at 4227 \AA, the shapes of the fractional linear polarization ($Q/I$) profiles were practically unknown 
\citep[e.g.,][]{trujillo_2_bruckner63,trujillo_2_stenflo74,trujillo_2_wiehr75}. 
In 1980 \cite*{trujillo_2_stenflos80} published the results of a first systematic exploration of the linearly polarized spectrum produced by scattering processes in the solar atmosphere. Of particular interest was finding  that the $Q/I$ pattern across the D-lines of Na {\sc i} changes sign between the D$_2$ and D$_1$ lines\footnote{This paper follows the usual convention of choosing the reference direction for $Q>0$ along the parallel to the observed solar limb.}. Even more interesting was  discovering the first hints of a similar sign reversal in the $Q/I$ pattern of the K and H lines of Ca {\sc ii}, which are 35 \AA\ apart. In a theoretical paper \cite{trujillo_2_stenfloHK80} could show that this is due to quantum mechanical interference between the two upper atomic states with total angular momentum $J=3/2$ and $J=1/2$.

It was however in 1983 when the first surveys of scattering line polarization were published. The observations of the UV spectral region between 3165 \AA\ and 4230 \AA\ were obtained using the vertical spectrograph of the Kitt Peak McMath Telescope 
\citep*[]{trujillo_2_stenflo83a}, while the survey between 4200 and 9950 \AA\ was achieved with the McMath Telescope using the Fast Fourier Spectrometer (FTS) as a polarimeter \citep[]{trujillo_2_stenflo83b}. Although the typical spectral resolution was 0.1 \AA\ (for the UV region) and the polarimetric sensitivity ${\sim}10^{-3}$, such observational surveys revealed a number of interesting linear polarization signals, most of which were considered as enigmatic. In particular, it was found that many spectral lines for which $\epsilon_Q{\approx}0$ (with $\epsilon_Q$ the contribution of scattering processes to the emissivity in Stokes $Q$) 
show instead conspicuous linear polarization peaks (e.g., the 8662 \AA\ line of Ca {\sc ii} whose $\epsilon_Q=0$ showed a mysterious, positive polarization profile, which was considered to provide a challenge for the theoretical efforts). Another unexpected finding was that the CN molecule shows significant linear polarization, increasing to a maximum at each band head. The impossibility of explaining the observed enigmatic $Q/I$ features via the standard theory of scattering line polarization led \cite{trujillo_2_stenflo83b} to write, ``clearly, the theoretical developments lag far behind in providing answers to the questions posed by our data".

It is interesting to mention that in the same year 
\cite{trujillo_2_landi83} published his paper on `Polarization in Spectral Lines:  A Unifying Theoretical Approach', where the scattering line polarization phenomenon is described as the temporal succession of 1st-order absorption and re-emission processes, interpreted as statistically independent events (complete redistribution in frequency).  
Actually, the phenomenon of scattering polarization in a spectral line is intrinsically a 2nd-order process \citep[e.g., the review by][]{{trujillo_2_casini-landi07}}, where frequency correlations between the incoming and outgoing photons can occur (partial redistribution in frequency). However, it is still possible to treat consistently the phenomenon of scattering to 1st order if the atomic system is illuminated by a spectrally flat radiation field\footnote{For the flat-spectrum approximation to hold, the incident  radiation field must be flat over a frequency interval $\Delta{\nu}$ larger than the natural width of the atomic levels, and, when coherences between non-degenerate levels are involved, $\Delta{\nu}$ must then be larger than the corresponding Bohr frequencies \citep[]{trujillo_2_landibook}.}. The key point to keep in mind is that within the framework of the 1st-order theory of spectral line polarization, the solution of the statistical equilibrium equations for the multipolar components of the atomic density matrix determines the excitation state of the atomic or molecular system, from which the emergent radiation can then be calculated by solving the radiative transfer equations \citep[see][for the numerical methods and computers programs with which we could solve this {\em Non-LTE Problem of the Second Kind}]{trujillo_2_trujillo99,trujillo_2_trujillo-tubin,trujillo_2_manso-bueno-spw3}. 

The 1983 exploratory surveys provided   
``a glimpse of the wealth of information accessible to us in this new field of observational solar physics" and made obvious the interest in pursuing the development of the Z\"urich Imaging Polarimeter (ZIMPOL), an instrument capable of measuring the polarization of the solar spectrum with a sensitivity limited only by photon statistics. ZIMPOL was invented by \cite{trujillo_2_povel95}, a physicist working on the development of instrumentation at the ETH. Since then, it has been used by Stenflo and collaborators in many observing campaigns in combination with several solar telescopes, mainly with the 1.5m McMath Telescope at Kitt Peak. It was in 1997 when we could see in detail the beauty of the second solar spectrum at a polarimetric sensitivity of $10^{-5}$ in $Q/I$ \citep[see][]{trujillo_2_stenflo-keller97}. A variety of fascinating $Q/I$ signals were discovered, many of which were considered (again) as ``enigmatic" by Stenflo and collaborators \citep*[see also][]{trujillo_2_stenflos00}. A suitable way to appreciate the improvement achieved with ZIMPOL is to compare Wiehr's (1975) observation of the $Q/I$ pattern across the sodium doublet (see his Fig. 1c) with Fig. 8 of \cite{trujillo_2_stenflos80}, and both of them with Fig. 2$a$ of \cite{trujillo_2_stenflo-keller97}. 

The $Q/I$ observations of Stenflo and coworkers 
\citep[see also][]{trujillo_2_gandorfer00,trujillo_2_gandorfer02} were soon confirmed and extended to the full Stokes vector by several groups using the Canary Islands telescopes.  
For example, Fig. 4 of  \cite{trujillo_2_trujillos01}
shows an {\em on-disk} observation of the O {\sc i} triplet around 7774 \AA, which led to the discovery of {\em negative} $Q/I$ polarization for the weakest line at 7776 \AA. Another interesting observational result was that the D-lines of Na {\sc i} show anomalous Stokes $V/I_{\rm c}$ profiles in ``quiet" regions close to the North solar limb, with the red lobe much more enhanced than the blue one \citep[see Figs. 2 and 3 of][]{trujillo_2_trujillos01}.

\section{The Key Physical Mechanism: ``Zero-field" Dichroism}

As we have seen, the development of
ZIMPOL and other polarimeters allowed us to see the 
linearly polarized solar-limb spectrum (that is, the second solar spectrum) with an unprecedented
degree of polarimetric sensitivity, sufficient to 
be able to confirm that a variety of atomic and molecular lines show indeed surprising $Q/I$ features. Some of them, like the sign reversals in the $Q/I$ pattern of the Na {\sc i} D-lines or the three-peak structure of the observed $Q/I$ in the Ba {\sc ii} D$_2$ line, could be qualitatively explained on the basis of the standard theory of scattering line polarization, which assumes that the emergent linear polarization originates only from the emission term in the radiative transfer equation \citep[e.g.,][]{trujillo_2_stenflo97}. However, many of the observed $Q/I$ signals were so perplexing, in particular those detected in spectral lines for which the standard theory predicted zero or negligible scattering polarization, that they were considered as a true enigma and a challenge for the theorists.  

A few months after \cite{trujillo_2_stenflo-keller97} reported on their ZIMPOL observations, \cite{trujillo_2_trujillo-landi97} pointed out that the presence of population imbalances among the lower-level substates of the enigmatic line transitions (that is, the presence of lower-level atomic polarization) would produce an important contribution to the emergent $Q/I$ through the ensuing {\em differential absorption of polarization components} (i.e., ``zero-field" dichroism). In order to show this, \cite{trujillo_2_trujillo-landi97} chose a line transition with $J_l=1$ and $J_u=0$, which according to the standard theory of scattering line polarization should be intrinsically unpolarizable because its upper level cannot be polarized so that $\epsilon_Q=0$ (i.e., a ``null line''). They formulated the problem by applying the density-matrix theory for the generation and transfer of polarized radiation and solved the resulting system of non-linear equations via the iterative method outlined in \cite{trujillo_2_trujillo99}. In this way, they could
show that the radiation field's anisotropy in solar-like atmospheres induces significant population imbalances among the sublevels of the lower level, even in the presence of the typical rates of elastic and inelastic collisions in the solar atmosphere. They concluded that such a lower-level atomic polarization can give rise to significant $Q/I$ amplitudes through selective absorption of polarization components (i.e., through the absorption term, $\eta_QI$, of the radiative transfer equation).

The radiative transfer equations of the scattering line polarization problem \citep[e.g., see Eqs. (7) and (8) in][]{trujillo_2_trujillo99}
indicate that the emergent spectral line polarization produced by the presence of atomic level polarization has, in general,
two contributions: one due to selective emission of polarization components (caused by the population imbalances of the upper level)
and an extra one due to selective absorption of polarization components (caused by the population imbalances of the lower level). Probably, the best way to understand this important 
fact is via the following generalization of the Eddington-Barbier formula \citep[see][]{trujillo_2_trujillo99,trujillo_2_trujillo-spw3}, which establishes that the emergent $Q/I$ at the line center of a sufficiently strong spectral line when observing along a line of sight (LOS) specified by $\mu={\rm cos}{\theta}$ (with $\theta$ the heliocentric angle) is approximately given by

\begin{equation}
Q/I\,\approx\,\epsilon_Q/\epsilon_I\,-\,\eta_Q/\eta_I\,{\approx}\,{3\over{2\sqrt{2}}}(1-\mu^2)
[{\cal W}\,\sigma^2_0({J_u})\,-\,{\cal Z}\,\sigma^2_0({J_l})],
\end{equation} 
where ${\cal W}$ and ${\cal Z}$ are numerical factors that depend on the quantum numbers of the transition
(e.g., ${\cal W}=0$ and ${\cal Z}=1$ for a transition with $J_l=1$ and $J_u=0$), while $\sigma^2_0=\rho^2_0/\rho^0_0$ quantifies the 
{\em fractional atomic alignment} of the upper or lower level of the line transition under consideration\footnote{For example, $\rho^0_0(J=1)=(N_1+N_0+N_{-1})/\sqrt{3}$ and $\rho^2_0(J=1)=(N_1-2N_0+N_{-1})/\sqrt{6}$, where $N_1$, $N_0$ and $N_{-1}$ are the populations of the magnetic sublevels.}. Note that in Eq. (1) the $\sigma^2_0$ values are those corresponding to the atmospheric height where the line-center optical depth is unity along the LOS. The first term of Eq. (1) is due to {\em selective emission} of polarization components, while the second term is caused by {\em selective absorption} of polarization components (i.e., ``zero-field" dichroism). An important point to emphasize here is that the presence of lower-level atomic polarization may have a significant feedback on the polarization of the upper levels \citep[see Eqs. (31) and (32) of][]{trujillo_2_trujillo01}, so that a calculation of the emergent $Q/I$ ignoring the selective absorption contribution (i.e., neglecting the second term of Eq. (1), as is justified for an optically thin plasma) will in general give a $Q/I$ amplitude different from that corresponding to the unpolarized lower-level case. In summary, the presence of atomic polarization in the lower level of a given spectral line may have two possible impacts on the emergent $Q/I$: one due to the modification of $\epsilon_Q$ due to the above-mentioned feedback effect on $\sigma^2_0(J_u)$, and an extra one coming from the selective absorption of polarization components. Interestingly, off-limb observations of a ``null line'' would tend to show zero linear polarization (because for such LOS we approach the optically thin limit), while on-disk observations can show conspicuous $Q/I$ signals due to ``zero-field" dichroism \citep[]{trujillo_2_trujillo-nature02}.

\begin{figure}
\begin{center}
\includegraphics[width=9.0 cm]{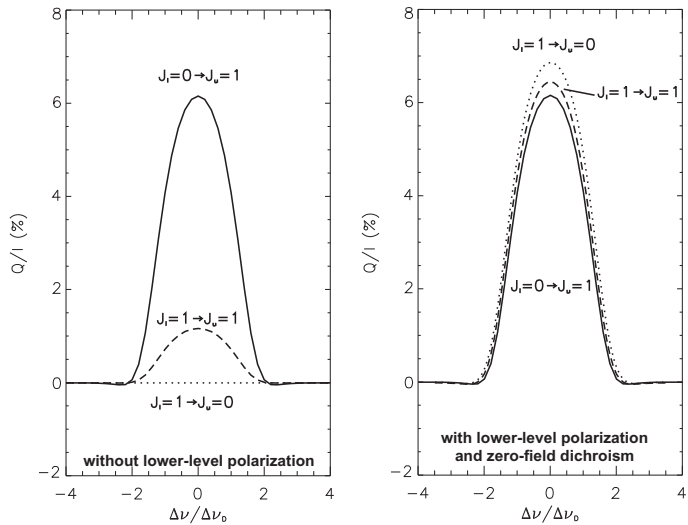}
\end{center}
\caption{The emergent $Q/I$ profiles ($\mu=0.1$) of the indicated three line transitions calculated in a model atmosphere with $T=6000$ K. Left panel: assuming that the lower level is unpolarized. Right panel: taking into account the full impact of lower-level polarization. Note that the linear polarization of the ``null line" (i.e., that with $\epsilon_Q=0$ because $J_u=0$) is due only to ``zero-field" dichroism. From \cite{trujillo_2_trujillo99}.}
\label{trujillo_2_fig:dichroism}
\end{figure}

It is important to note that there are two key mechanisms capable of producing {\em directly} atomic level polarization through the absorption of radiation \citep[e.g.,][]{trujillo_2_happer72,trujillo_2_trujillo01}: {\em upper-level} selective population pumping (which occurs when some substates of the upper level have more chances of being populated than others) and {\em lower-level} selective depopulation pumping (which occurs when some substates of the lower level have more chances of being depopulated than others). For this to occur the pumping light must be {\em anisotropic} and/or {\em polarized} and/or to have {\em spectral structure} over a frequency interval $\Delta{\nu}$ smaller than the frequency separation between the sublevels (e.g., when it is {\em monochromatic}). Typically, in solar-like atmospheres the pumping light is unpolarized, anisotropic and broad-band (i.e., all the allowed radiative transitions in the atom under consideration are simultaneously excited), while in an optical pumping experiment with a laser the incident light is polarized, anisotropic and monochromatic. In addition to the above-mentioned mechanisms that allow the direct transfer of ``order" from the radiation field to the atomic system, we have to take into account the so-called {\em repopulation pumping} mechanism. This pumping occurs either when the lower level is repopulated as a result of the spontaneous decay of a {\em polarized} excited state or when the upper level is repopulated as a result of absorptions from a {\em polarized} lower level. Obviosly, {\em lower-level} selective depopulation pumping is the only mechanism that can produce atomic polarization in the lower level of a resonance line transition with $J_l=1$ and $J_u=0$.

Consider the three line transitions of Fig. 1, and the corresponding
emergent $Q/I$ profiles obtained by solving numerically the scattering polarization problem in an unmagnetized model of the solar atmosphere assuming a two-level atomic model for each line independently. The left panel of Fig. 1 corresponds to calculations carried out assuming that the lower level is completely unpolarized, while the right panel takes into account the full impact of lower-level polarization. Note that when the atomic polarization of the lower level is taken into account then the ``null line'' (i.e., that with $J_{l}=1$ and $J_{u}=0$) shows the largest 
$Q/I$ amplitude. Isn't it fascinanting? I mean the fact that ``zero-field" dichroism (that is, differential absorption of polarization components) is a very efficient mechanism for producing linear polarization in the spectral lines that originate in a stellar atmosphere.

\section{Evidence for ``Zero-field" Dichroism in the Solar Atmosphere}

In their theoretical investigation \cite{trujillo_2_trujillo-landi97} 
pointed out that ``zero-field" dichroism is probably the physical mechanism that produces the enigmatic linear polarization observed in a variety of spectral lines of the second solar spectrum, but did not demonstrate that this is actually the case for particular spectral lines whose observed $Q/I$ was considered enigmatic. The first evidence that zero-field dichroism is indeed at work in the quiet solar atmosphere was achieved through the Mg {\sc i} $b$ lines \citep[see][]{trujillo_2_trujillo99,trujillo_2_trujillo01}, but let us review also some of the other 
theoretical developments that took place during those years
because this will help clarify the underlying physics and diagnostic possibilities of the enigmatic $Q/I$ features.

\subsection{The Enigmatic Na {\sc i} D-lines}

\begin{quote}
{\small ``{\em The observed polarization peaks are an enigma, 
a challenge for the theorists.}" \citep{trujillo_2_stenflos00}}. 

{\small ``{\em Here I report a mechanism that may explain these observations, in which it is assumed that the populations of the electronic ground state of the sodium atom are not equal.}" \citep{trujillo_2_landi98}.}
\end{quote}

The fractional linear polarization of the sodium doublet
observed by \cite{trujillo_2_stenflo-keller97} can be seen in the solid line of Fig. 2a. It shows a {\em symmetric} $Q/I$ profile at the very line center of D$_1$ whose amplitude is only 3.5 times smaller than that corresponding to the D$_2$ central peak. \cite{trujillo_2_stenflo-keller97} considered the D$_1$ peak as enigmatic because the D$_1$ line is the result of a transition with $J_l=J_u=1/2$, which should be in principle intrinsically unpolarizable because for a spectral line with such quantum numbers there is no contribution of atomic level alignment to $\epsilon_Q$ and $\eta_Q$. 

\begin{figure}
\begin{center}
\includegraphics[width=9.0 cm]{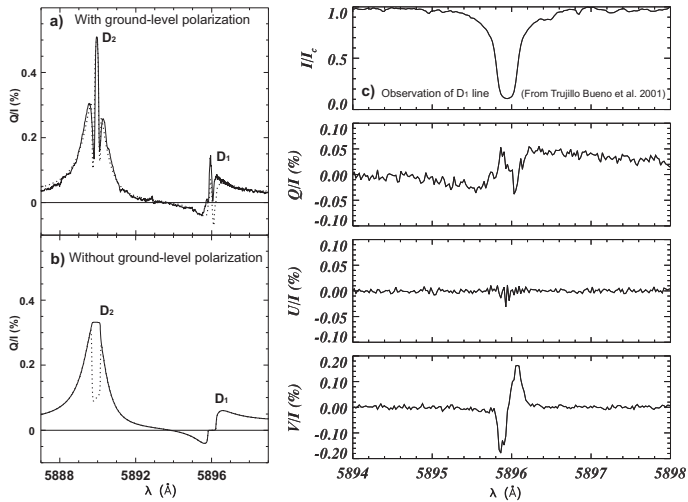}
\end{center}
\caption{The solid line of panel {\bf (a)} shows the $Q/I$ observed by \cite{trujillo_2_stenflo-keller97} at $\mu=0.05$. The dotted line is Landi Degl'Innocenti's (1998) parameterized fit accounting for HFS and assuming $\sigma^2_0(F_l=1)=0.0125$ and $\sigma^2_0(F_l=2)=0.0238$ at the top of the solar atmosphere. The dotted line of panel {\bf (b)} shows the calculated $Q/I$ when the ground-level is assumed to be unpolarized, while the solid line indicates the emergent $Q/I$ when in addition the HFS of sodium is neglected. Panel {\bf (c)} shows the full Stokes-vector observations of Trujillo Bueno et al. (2001) at $\mu=0.1$.}
\label{trujillo_2_fig:landi}
\end{figure}

\cite{trujillo_2_landi98} tried to provide an explanation 
of the $Q/I$ pattern observed by \cite{trujillo_2_stenflo-keller97} in the Na {\sc i} D-lines. To this end, he considered the $B=0$ G case within the framework of a heuristic partial redistribution theory of line scattering polarization, but assuming frequency-coherent scattering and constant anisotropy within each line. He applied a parameterized modeling approach based on his analytical expressions of $\epsilon_X$ and $\eta_X$ (with $X=I,Q$) for a multiplet with hyperfine structure (HFS), and accounting for the possibility of ground-level polarization. The key free parameters of his modeling were the values of the fractional atomic alignment of the two lower HFS levels of the D-lines (that is, the values of $\sigma^2_0(F_l=1)$ and $\sigma^2_0(F_l=2)$, where $\sigma^2_0(F)=\rho^2_0(F)/\rho^0_0(F)$ is the fractional atomic alignment of the hyperfine $F$-level under consideration). \cite{trujillo_2_landi98} increased and adjusted these and a few extra free parameters till obtaining a best fit to the observations.  
The resulting fit was so striking (see the dotted line of Fig. 2a) that he concluded that the ground level of Na {\sc i} must be polarized in the region of the solar atmosphere where the D$_1$ line polarization originates. As clarified by \cite{trujillo_2_landi99}, the emergent $Q/I$ was found to be fully dominated by the $\epsilon_Q$ term of the Stokes-vector transfer equation (i.e., in his parameterized modeling the contribution of ``zero-field" dichroism turned out to be insignificant). Note also that in the presence of the imposed ground level polarization (which was needed to get atomic alignment in the upper $F$-levels of D$_1$) the three-peak structure of the $Q/I$ profile of the D$_2$ line and the relative amplitudes of the D$_2$ and D$_1$ lines are in good agreement with the observations. However, the theoretical $Q/I$ profile around the line center of the D$_1$ line is antisymmetric (see the dotted line of Fig. 2a), in contrast with the symmetric $Q/I$ profile observed by \cite{trujillo_2_stenflo-keller97} (see the solid line of Fig. 2a), but in agreement with the observations of Trujillo Bueno et al. (2001) shown in Fig. 2c.

The investigation by \cite{trujillo_2_landi98}
was based on formulae derived for the unmagnetized reference case. His conclusion that the magnetic field of the lower solar chromosphere must be either isotropically distributed and extremely weak (with $B < 0.01$ G) or, alternatively, practically radially oriented was based on ($a$) the substantial amount of ground-level polarization required to fit the $Q/I$ line-center amplitudes of the D$_2$ and D$_1$ lines and ($b$) the assumption that the atomic polarization of the ground level of Na {\sc i} must be sensitive to much weaker fields than the atomic polarization of the upper level of the D$_1$-line. Today, we have a better knowledge of the magnetic sensitivity of the atomic polarization of the sodium levels, but this is a subject we will discuss in \S~4.3. For the moment let us continue with the $B=0$ G case and emphasize that in Landi Degl'Innocenti's (1998) modeling the requirement of {\em a substantial amount of ground level polarization} is essential for explaining the line-center $Q/I$ peaks observed in both, the D$_1$ and D$_2$ lines. 

A very important question is whether or not the anisotropy of the radiation field in the real solar atmosphere is high enough so as to be able to lead to the sizable values of $\sigma^2_0(F_l=1)$ and $\sigma^2_0(F_l=2)$ that \cite{trujillo_2_landi98} had to choose ad hoc for fitting the $Q/I$ observations (see such $\sigma^2_0(F)$ values in the caption of Fig. 2). Are they consistent for the $B=0$ G case? Let us point out now what we know from self-consistent solutions of the statistical equilibrium equations for a Na {\sc i} model atom with the levels $3^2S_{1/2}$, $3^2P_{1/2}$ and $3^2P_{3/2}$, as formulated within the framework of the density-matrix theory of spectral line polarization described in the book by \cite{trujillo_2_landibook}. 

In order to induce the ground-level polarization values given in the caption of Fig. 2 (i.e., $\sigma^2_0(F_l=2)\,{\approx}\,0.02$) we would need a spectral line radiation with an anisotropy factor $w=\sqrt{2}J^2_0/J^0_0\,{\sim}\,0.1$ at the height in the solar atmosphere where the line-center optical depth is unity along the LOS \citep[see in Fig. 1 of][that at $B=0$ gauss $\sigma^2_0(F_l=2)\,{\approx}\,0.024$ for $w\,{\approx}\,0.12$]{trujillo_2_trujillos02}. Do we have this level of anisotropy around 1000 km in the ``quiet" solar atmosphere? Fig. 3 shows the height variation of the anisotropy factor of the D-line radiation calculated in the hydrodynamical simulations of internetwork chromospheric dynamics by \cite{trujillo_2_carlsson-stein02}. While the solid line of Fig. 3 indicates the temporally averaged anisotropy factor at each atmospheric height (e.g., we have ${\langle w \rangle}=0.07$ at $h=1000$ km), the gray-shaded area shows that the range of variation for all time steps of the simulation is very important, since $w$ varies  between zero and 0.15. The amplitude of the calculated $Q/I$ for a LOS with $\mu=0.1$ varies between zero and 0.12\%, while $\langle Q/I \rangle\,{\approx}\,0.05\%$ \citep[]{trujillo_2_aar-jtb-09}. The results of our radiative transfer investigation of the scattering polarization of the Na {\sc i} D-lines in such dynamical models of the solar chromosphere show that in the absence of magnetic and/or collisional depolarization the amplitude of the emergent $Q/I$ in the D$_1$ line can be of the same order of magnitude as that of the observed profiles. This is reinforced by the conclusion of \cite{trujillo_2_asensios03} that below 700 km the kinetic temperature of the solar atmospheric plasma should be even lower than that corresponding to the cool phases of Carlsson \& Stein's (2002) simulations, since this could increase further the anisotropy of the radiation field through the modified  
temperature gradients \citep[see Fig. 4 in][]{trujillo_2_trujillo01}. 
Moreover, in our radiative transfer modeling of the D$_1$ line ``zero-field" dichroism does make a significant contribution to the emergent $Q/I$! 

\begin{figure}
\begin{center}
\includegraphics[width=9.5 cm]{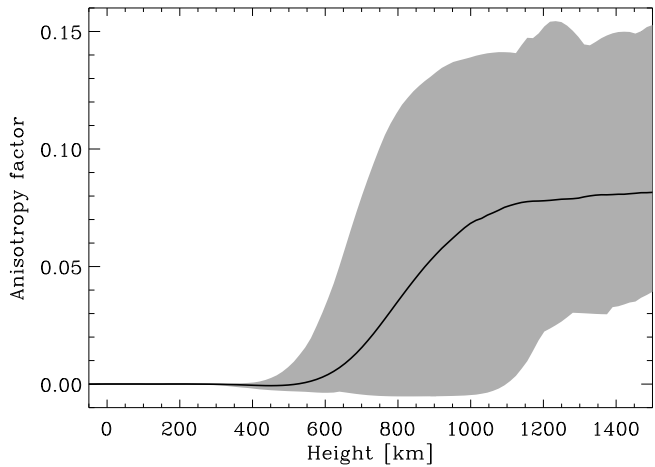}
\end{center}
\caption{Range of variation of the anisotropy factor of the D-line radiation in the hydrodynamical simulations of internetwork chromospheric dynamics by \cite{trujillo_2_carlsson-stein02}. From \cite{trujillo_2_aar-jtb-09}.}
\label{trujillo_2_fig:anisotropy}
\end{figure}

Although the previously advanced result is very encouraging, the shape of the calculated $Q/I$ of the D$_1$ line is still antisymmetric, in sharp contrast with the symmetric profile observed by \cite{trujillo_2_stenflo-keller97}, but in agreement with the observations of Trujillo Bueno et al. (2001). 
Another discrepant point but of minor consequences is that $\sigma^2_0(F_l=2)/\sigma^2_0(F_l=1)=1.9$ in Landi Degl'Innocenti's (1998) modeling, while that resulting from a self-consistent calculation is $\sigma^2_0(F_l=2)/\sigma^2_0(F_l=1)=4$ for the unmagnetized case. 

\subsection{The Mg {\sc i} $b$-lines: Three Lines with the Same Upper Level}

\begin{quote}
{\small ``{\em It is impossible to come remotely close to a qualitative fit to the observed relative polarization amplitudes.}" \citep[]{trujillo_2_stenflos00}.}

{\small ``{\em The only way I see for increasing the emitted polarization in the Mg $b_1$ and $b_2$ lines, so as to bring it to the same level of that corresponding to the Mg $b_4$ line (which has $J_l=0$), is via the dichroism contribution.}" \citep{trujillo_2_trujillo99}.}
\end{quote}

The observations by \cite{trujillo_2_stenflos00} showed similar $Q/I$ amplitudes for the three Mg {\sc i} $b$-lines, which share the same upper level, $4^3{\rm S}_1$, whose $J_u=1$. We point out that in Eq. (1) ${\cal W}=1$ and ${\cal Z}=0$ for the $b_4$ line at 5167 \AA\  (whose $J_l=0$), ${\cal W}={\cal Z}=-1/2$ for the $b_2$ line at 5173 \AA\ (whose $J_l=1$), and ${\cal W}=0.1$ and ${\cal Z}{\approx}0.6$ for the $b_1$ line at 5184 \AA\ (whose $J_l=2$). First, assume that there is no atomic polarization in the lower levels ($3^3{\rm P}_{0,1,2}$) of the Mg {\sc i} $b$-lines. We would have then $Q/I\,{\approx}\,{\cal W}{\sigma}^2_0({J_{u}}=1)$, with each $b$-line having its corresponding ${\sigma}^2_0({J_{u}}=1)$ value at the atmospheric height ($h$) where the line-center optical depth is unity along the observed LOS, whose $\mu=0.1$. Therefore, in order to be able to explain such observations assuming that the lower levels are unpolarized, ${\sigma}^2_0({J_{u}}=1)$ would have to change from having the value $X$ at $h(b_4){\approx}827$ km (with $X>0$ and similar to the $Q/I$ amplitude observed in the $b_4$ line), to becoming negative (i.e., $-2X$) at $h(b_2){\approx}952$ km, and then positive again (and similar to $10X$ !) at $h(b_1){\approx}1013$ km. This peculiar variation of ${\sigma}^2_0({J_{u}}=1)$ seems very unlikely, even when the optical pumping processes resulting from the three near-IR transitions between the $4^3{\rm S}_1$ level and the $4^3{\rm P}_{0,1,2}$ levels 
are taken into account. When only radiative pumping processes in the Mg {\sc i} $b$-lines are considered, assuming still that their lower metastable levels are unpolarized, one finds ${\sigma}^2_0({J_{u}}=1)\,{\approx}\,[{\cal A}(b_4)-{{3}\over{2}}\,{\cal A}(b_2)+{{1}\over{2}}\,{\cal A}(b_1)]/9$, where ${\cal A}=J^2_0/J^0_0$ is the ``degree of anisotropy" of the radiation field in each of the indicated spectral lines. Note that if ${\cal A}(b_4)={\cal A}(b_2)={\cal A}(b_1)$ then $Q/I=0$ in the three lines. Although this is {\em not} the case in semi-empirical models of the solar atmosphere (i.e., in reality ${\cal A}(b_4){>}{\cal A}(b_2){>}{\cal A}(b_1){>}0$ at any atmospheric height between $h(b_4)$ and $h(b_1)$), there is no possibility of ending up with the above-mentioned peculiar variation for ${\sigma}^2_0({J_{u}}=1)$ through anisotropic radiation pumping in the Mg {\sc i} $b$-lines. Assume now that the lower levels are polarized, so that we have to include also the ``zero-field" dichroism term of Eq. (1). As pointed out by \cite{trujillo_2_trujillo99}, the observations could then be easily explained if 
${\sigma}^2_0({J_{l}}=1){\approx}3X$ at $h(b_2)$ and ${\sigma}^2_0({J_{l}}=2){\approx}-2X$ at $h(b_1)$. It is amazing that these are precisely the fractional atomic polarization values that result when doing self-consistent scattering polarization calculations for a 19-level model atom in a semi-empirical model of the solar atmosphere \citep[see Fig. 8 of][]{trujillo_2_trujillo01}. A similar explanation applies to the three lines of Ca {\sc i} at 6103, 6122 and 6162 \AA.

\subsection{The Ca {\sc ii} IR Triplet}

\begin{quote}
{\small ``{\em The Ba {\sc ii} 6497 \AA\ line, which like the Ca {\sc ii} 8662 \AA\ line should be intrinsically unpolarizable, 
exhibits a strong and well defined polarization peak. This result further underscores that we are dealing with a fundamental problem, for which we lack a physical explanation.}" \citep[]{trujillo_2_stenflos00}.}
\end{quote}

The reason why according to \cite{trujillo_2_stenflos00} the Ca {\sc ii} 8662 \AA\ line should be intrinsically unpolarizable is because its upper level has  $J_u=1/2$, so that ${\cal W}=0$ in Eq. (1) and $\epsilon_Q=0$. Since calcium has no hyperfine structure it is clear from Eq. (1) that the {\em only} possibility to explain the $Q/I$ observed in the Ca {\sc ii} 8662 \AA\ line is through the second term of this equation, which accounts for the contribution of  ``zero-field" dichroism. \cite{trujillo_2_manso-bueno-prl} demonstrated quantitatively that this is actually the case in the solar chromosphere (see their Fig. 3; see also Fig. 5 below). To this end, they solved the problem of the generation and transfer of polarized radiation by taking fully into account all the relevant optical pumping mechanisms in multilevel atomic models (see a description of the computer program in Manso Sainz \& Trujillo Bueno 2003a). They finished their letter in Physical Review by pointing out that ``{\em zero-field dichroism may also be operating in other astrophysical objects (e.g., accreting systems) and should be fully taken into account when interpreting spectropolarimetric observations in other spectral lines besides the Ca {\sc ii} IR triplet itself, whose polarization has been observed recently in supernovae}". It is particularly gratifying that this new polarization mechanism that \cite{trujillo_2_trujillo-landi97}  and \cite{trujillo_2_trujillo99} pointed out considering a particularly simple spectral line model has allowed us to understand many of the enigmatic features of the second solar spectrum, while at the same time helping other researchers to interpret their non-solar observations \citep[e.g.,][]{trujillo_2_kuhn07}.  

\subsection{The Oxygen IR Triplet: Three Lines with the Same Lower Level}

Another fascinating group of lines is that comprising the three lines of the O {\sc i} triplet around 7774 \AA, which share the same lower level whose $J_l=2$. When observed on the solar disk, the line-center radiation in these transitions comes from the solar photosphere. In fact, the height in a realistic atmospheric model where the line-center optical depth is unity along a LOS with $\mu=0.1$ lies between 250 and 300 km, approximately. Interestingly, the theoretical prediction for the relative amplitudes of the emergent $Q/I$ in such oxygen lines is shown in the left panel of Fig. 4, which corresponds to a simulated on-disk observation very close to the solar limb. As shown by the solid and dashed lines of the figure, the presence of lower-level polarization and the ensuing differential absorption of polarization components (``zero-field" dichroism) gives rise to a negative $Q/I$ signal for the 7776 \AA\ line, while the theoretical prediction for the other two lines is that both of them should show positive $Q/I$ signals, but with a larger amplitude for the 7774 \AA\ line. On the contrary, in the absence of lower-level polarization the emergent $Q/I$ in the 7776 \AA\ line should show a tiny positive signal, while the amplitude of the other two lines should be similar when observing on the solar disk (see the dotted line of Fig. 4).

\begin{figure}
\begin{center}
\includegraphics[width=13 cm]{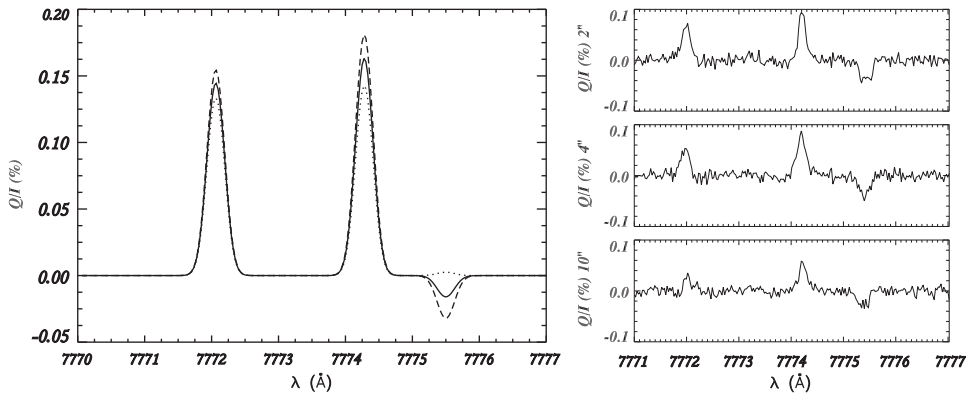}
\end{center}
\caption{Left panel: Close to the limb simulation of the $Q/I$ signals in the O {\sc i} triplet around 7774 \AA, taking into account the contribution of ``zero-field" dichroism for the following three values of the depolarization rate $D$ of elastic collisions with neutral hydrogen atoms: $D=0$ (dashed line), $D=10^6\,{\rm s}^{-1}$ (solid line) and $D=10^{18}\,{\rm s}^{-1}$ (dotted line, which shows the unpolarized lower-level case). Right panel: the TH\'EMIS observations of \cite{trujillo_2_trujillos01} for three distances from the ``quiet" Sun limb.} 
\label{trujillo_2_fig:oxi-modeling}
\end{figure}

A direct comparison of the radiative transfer modeling predictions 
of the left panel of Fig. 4 with the spectropolarimetric observations of 
\citet{trujillo_2_trujillos01} (see the right panel of Fig. 4) suggests that {\em zero-field dichroism is operating even in the solar photosphere}.

\section{The Magnetic Sensitivity of the Solar Spectrum}

\begin{quote}
{\small ``{\em I think I can safely say that nobody understands quantum mechanics}" \\(R. P. Feynman, The Character of Physical Law, 1965, MIT Press, p. 129)}
\end{quote}

The magnetic sensitivity of the solar spectrum is due to the Zeeman effect (that is, to the wavelength shifts between the $\pi$ and $\sigma$ components caused by the splittings of the atomic energy levels)
and to a variety of less familiar physical mechanisms by means of which a magnetic field can create and destroy spectral line polarization. These unfamiliar mechanisms will be referred to here by the term ``Hanle effect", since they have to do with the various subtle ways in which a magnetic field can modify the atomic level  polarization created by anisotropic radiative pumping processes.

The Zeeman effect of a spatially resolved field can 
dominate the polarization of the emergent radiation if the splitting among the magnetic sublevels is a significant fraction of the spectral line width (which is much larger than the natural width of the atomic levels!). Typically, 100 G or more are needed to be able to observe the signature of the transverse Zeeman effect on the Stokes $Q$ and $U$ profiles, while much weaker resolved fields are enough to produce measurable Stokes $V$ amplitudes via the longitudinal Zeeman effect. The polarization of the Zeeman effect as a diagnostic tool is ``blind" to magnetic fields that are tangled on scales too small to be resolved, a disadvantage that does not apply neither to the Zeeman broadening of the Stokes $I$ profiles nor to the Hanle effect. The ``Hanle effect" modifies the population imbalances and the quantum coherences among the different magnetic sublevels, even among those pertaining to different $J$ (fine-structure) or $F$ (hyperfine-structure) levels. Let us see some examples of the remarkable effects that the atomic level polarization produces on the emergent spectral line polarization.

\subsection{The Hanle Effect in the Ca {\sc ii} IR Triplet}

It can be demonstrated that the atomic polarization originating from the magnetic sublevels pertaining to a given $J$ or $F$ level affects mainly the line center polarization, and that this polarization is significantly modified by the standard Hanle effect when the Zeeman splitting is of the same order of magnitude as the natural width of the level \citep[e.g.,][]{trujillo_2_landibook}. For such a modification to take place the magnetic field must be inclined with respect to the symmetry axis of the pumping radiation field. Approximately, the amplitude of the emergent spectral line polarization is sensitive to magnetic strengths between $0.1\,B_H$ and $10\,B_H$, where the critical Hanle field (in gauss) is

\begin{equation}
B_{\rm H}={\frac{1.137\times10^{-7}}{t_{\rm life}\,g_L}}\, ,
\end{equation}
with $g_L$ the level's Land\'e factor and 
$t_{\rm life}$ (in seconds) its radiative lifetime.
If the lower level of the line transition under consideration is the ground 
level or a metastable level, as happens with all the spectral lines considered in \S~3, its ${t_{\rm life}(J_l)}{\approx}1/B_{lu}J^0_0$ (with $J^0_0$ the mean intensity of the spectral line radiation), which for relatively strong spectral lines is typically between a factor $10^2$ and $10^3$ larger than the upper-level lifetime (${t_{\rm life}(J_u)}{\approx}1/A_{ul}$). For this reason, the lower-level Hanle effect is normally sensitive to magnetic fields in the milligauss range, while the upper-level Hanle effect is sensitive to fields in the gauss range. 

\begin{figure}
\begin{center}
\includegraphics[width=10 cm]{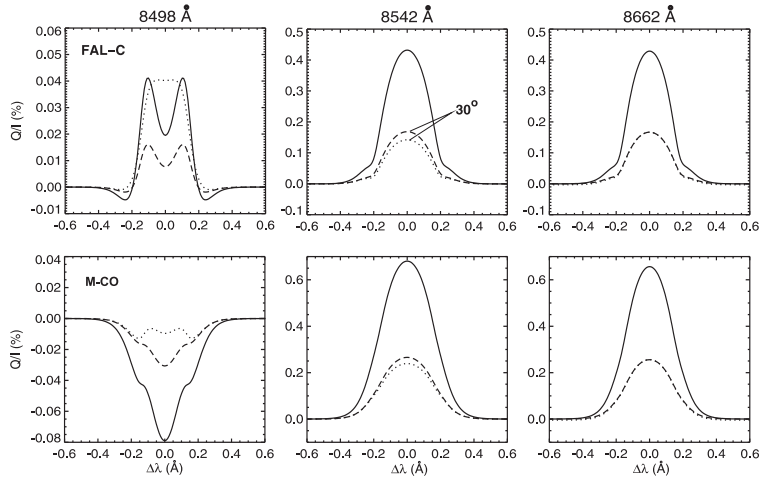}
\end{center}
\caption{$Q/I$ of the Ca {\sc ii} IR triplet calculated at $\mu=0.1$ in the ``hot" FAL-C model (upper panels) and in the ``cool" M-CO model (lower panels), assuming the presence of a magnetic field inclined by $30^{\circ}$ and with a uniformly distributed azimuth. Solid lines: $B\!=\!0$ G. Dotted lines: $B\!=\!0.3$ G. Dashed lines: $B\!=\!100$ G. Note that in the panels of the 8662 \AA\ line the dotted and dashed lines coincide, because for $B{>}0.1$ G its $Q/I$ is only sensitive to the orientation of the magnetic field. From \cite{trujillo_2_manso-bueno07}.}
\label{trujillo_2_fig:calcium}
\end{figure}

In this respect, a very suitable diagnostic window for investigating the thermal and magnetic structure of the ``quiet" solar chromosphere is the scattering polarization in the Ca\,{\sc ii} IR triplet (see Manso Sainz \& Trujillo Bueno 2003b, 2007). This can be seen also in Fig. 5, which shows the emergent $Q/I$ in two semi-empirical models of the solar chromosphere: the ``hot" FAL-C model of \cite*{trujillo_2_falc} (upper panels) and the ``cool" M-CO model of \cite{trujillo_2_avrett95}. Interestingly, although the magnetic sensitivity of the scattering polarization signals in the 8662 and 8542 \AA\ lines is dominated by the lower-level Hanle effect, the observations of \cite{trujillo_2_stenflos00} are compatible with the physical conditions of the ``hot" chromospheric model in the presence of significantly inclined fields (i.e., $\theta_B{\approx}30^{\circ}$) with a strength sensibly larger than 10 G (e.g., of the order of the 100 G required to saturate the Hanle effect of the upper level $P_{3/2}$). We point out, however, that fields of the order of 10 mG are required in order to be able to obtain with FAL-C the observed $Q/I$ amplitude of 0.04\% for the 8498 \AA\ line, which is the weakest of the Ca {\sc ii} IR triplet.

\begin{figure}
\begin{center}
\includegraphics[width=10 cm]{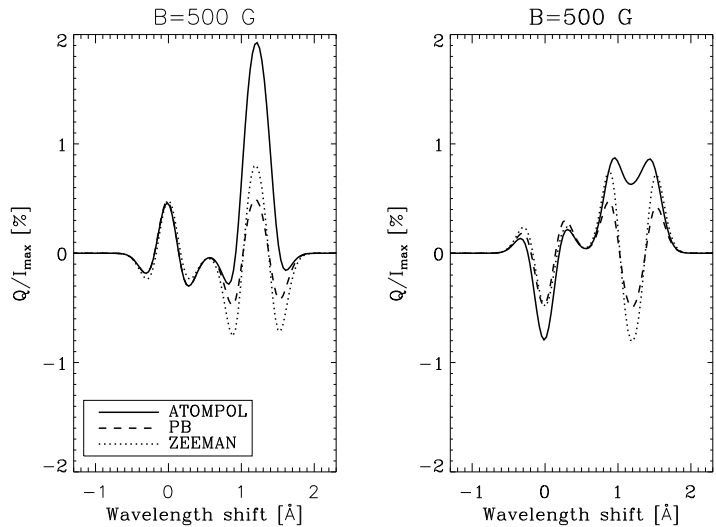}
\end{center}
\caption{The emergent Stokes $Q/I_{\rm max}$ profiles of the 
He {\sc i} 10830 \AA\ multiplet calculated for two scattering geometries: $90^{\circ}$ scattering (left panel) and forward scattering (right panel). Each panel shows the results of three possible calculations for the case of a 500 G horizontal magnetic field perpendicular to the LOS. The positive reference direction for Stokes $Q$ is along the direction of the horizontal magnetic field. For more information see \cite{trujillo_2_jtb-aar07}.}
\label{trujillo_2_fig:helium}
\end{figure}

\subsection{The Importance of the Paschen-Back Effect and the Physics of the He {\sc i} 10830 \AA\ Polarization}

In general, rigorous modeling of the spectral line polarization produced by the joint action of the Hanle and Zeeman effects requires calculating the wavelength positions and the strengths of the $\pi$ and $\sigma$ components within the framework of the Paschen-Back effect theory (e.g., Landi Degl'Innocenti \& Landolfi 2004). For the case of a horizontal magnetic field with $B=500$ G, Fig. 6 shows three possible calculations of the emergent linear polarization in the He {\sc i} 10830 \AA\ multiplet, both for $90^{\circ}$ scattering geometry (left panel) and for forward scattering geometry (right panel). The dashed and dotted lines neglect the influence of atomic level polarization. Their only difference is that the dotted line calculation assumed that the splitting between the magnetic sublevels of each $J$ level is linear with the magnetic strength (that is, the Zeeman effect regime), while the dashed line calculation took into account that the splitting produced by the magnetic field on each $J$ level is not necessarily small compared to the energy separation between the different $J-$levels of the corresponding $(L,S)$ term (that is, the Paschen-Back effect). The solid line calculation shows what happens when we additionally account for the contribution of the selective emission and selective absorption of polarization components caused by the presence of atomic level polarization. Clearly, for some spectral lines the impact of the presence of atomic level polarization on the emergent spectral line polarization can be very important, even in the presence of relatively strong fields.

The theory of the Paschen-Back effect in a hyperfine structured multiplet allows us to model the important level-crossing regime in which the energy eigenvectors are gradually evolving from the form ${\vert} J I F f \rangle$ (with $f$ the projection of the total angular momentum $F=J+I$ along the quantization axis) to the form ${\vert} J I M_J M_I \rangle$ as the magnetic field increases. This range between the limiting cases of ``weak'' fields (Zeeman effect regime) and ``strong'' fields (complete Paschen-Back regime) is called the incomplete Paschen-Back effect regime. The reason why it is so important for understanding the magnetic sensitivity of the solar spectrum is because the level crossings and repulsions that take place in this regime give rise to subtle modifications of the atomic level polarization and, therefore, to a number of remarkable effects on the emergent spectral line polarization \citep*[e.g.,][]{trujillo_2_bommier80,trujillo_2_landi82,trujillo_2_trujillos02,trujillo_2_belluzzis07}. Let us see some examples for the case of the D$_2$ and D$_1$ lines of sodium. 

\subsection{The Magnetic Sensitivity of the Na {\sc i} D-lines and 
the Physical Origin of their Enigmatic Polarization}

In the solar atmosphere the {\em depopulation pumping} mechanism discussed in \S~2 does {\em not} play any role on the ground level polarization of sodium\footnote{This contrasts with the case of optical pumping experiments with D$_1$-line excitation only, where it is possible to produce $F$-level polarization {\em directly} by the absorption of narrow-line D$_1$ light \citep[e.g.,][]{trujillo_2_franzen57}. For this reason, the results of the potassium experiment with D$_1$ laser light mentioned by J. O. Stenflo in this workshop do not come as a real surprise.}. The key mechanism is {\em repopulation pumping} \citep[]{trujillo_2_trujillos02,trujillo_2_casinis02}. First, only $J$ alignment can be created in the $P_{3/2}$ level via the D$_2$ broad-line excitation that we have in a stellar atmosphere. This electronic alignment of the level $P_{3/2}$ cannot be transferred to the ground level because its $J_l=1/2$. However, since the HFS of the level $P_{3/2}$ is of the same order of magnitude as its natural width, we can have a significant HFS interaction during the lifetime of the level $P_{3/2}$, with the result that the hyperfine coupling of the nucleus to the electrons can transform the $J$ alignment into $F=J+I$ alignment before the de-excitation process. This alignment of the $F$ levels of the $P_{3/2}$ term is transferred to the HFS levels of the $S_{1/2}$ ground term via spontaneous emission in the D$_2$ line, and then from the $S_{1/2}$ $F$-levels to those of the $P_{1/2}$ term via radiative absorptions in the D$_1$ line. This has a very important consequence, namely that in spite of the sizable differences between the lower and upper level lifetimes, the atomic polarization of the lower and upper $F$-levels of the D$_1$ line {\em are sensitive to the same magnetic field strengths}. Equally important is the conclusion that independently of the magnetic field inclination (i.e., even for a purely vertical field) the atomic alignment of the lower and upper levels of the D$_1$ line are suddenly reduced for $B>10$ G \citep[see Fig. 1 of][]{trujillo_2_trujillos02}. It can be shown analytically that the vanishing of atomic alignment in the levels with $J=1/2$ sets in when the electronic and nuclear angular momenta are decoupled for the atom in the excited state $P_{3/2}$ \citep[see][]{trujillo_2_casinis02}. This decoupling is reached in the limit of the complete Paschen-Back effect of the level $P_{3/2}$. Although for the case of Na {\sc i} this limit occurs for $B{\gtrsim}100$ G, the atomic polarization of the D$_1$ levels is practically negligible for $B{>}10$ G.

The magnetic sensitivity of the atomic polarization of the Na {\sc i} levels has the following important consequences for the emergent linear polarization in the D$_2$ and D$_1$ lines of Na {\sc i}.

{\bf (A) The D$_2$ Line}

For $B{<}10$ G the atomic polarization of the ground level of sodium can be very significant, especially if the magnetic field is nearly vertical or if it is sensibly weaker than $10^{-2}$ G \citep[see Fig. 1 of][]{trujillo_2_trujillos02}. The presence of ground-level polarization has a non-negligible feedback on the atomic polarization of the $P_{3/2}$ levels, which in turn produces a significant but small enhancement of the emergent linear polarization in the D$_2$ line core, with respect to the  completely unpolarized ground level case \citep[see the left panels of Fig. 2 in][]{trujillo_2_trujillos02}. The influence of dichroism on the D$_2$ line polarization is, however, negligible. 

For $B>10$ G the scattering polarization of the Na {\sc i} D$_2$ line is fully dominated by the atomic polarization of the $F$-levels of the $P_{3/2}$ upper term. Of great diagnostic interest is the enhancement of the line-center scattering polarization of the D$_2$ line of Na {\sc i} by a vertical magnetic field between 10 and 100 G \citep[see][]{trujillo_2_trujillos02}, which is due to the fact that as the incomplete Paschen-Back effect regime is reached the $\rho^2_0(F,F^{'})$ quantum interferences are modified because of the repulsions between the HFS magnetic sublevels having the same $f$ quantum number. Interestingly, this theoretical prediction was observationally confirmed by \cite{trujillo_2_stenflos02} via filter polarimetry of the solar atmosphere. They found that the scattering polarization in the sodium D$_2$ line shows an intermittent structure that can be explained in terms of magnetic enhancement of the scattering polarization in the network and/or  depolarization of the scattering polarization outside the network through the familiar Hanle effect of an inclined field\footnote{The sensitivity of the D$_2$ line of Na {\sc i} to the upper level Hanle effect lies between 0.5 G and 50 G, approximately.}.

In summary, the main physical origin of the $Q/I$ observed in the D$_2$ line of Na {\sc i} appears to be upper-level atomic polarization. If ground-level polarization does not play any crucial role in the scattering polarization of the D$_2$ line, then the three-peak structure of the observed $Q/I$ profile could perhaps be a consequence of the fact that the anisotropy of the pumping radiation depends on the wavelength within the D$_2$ line itself \citep*[cf.][]{trujillo_2_holzreuter05}. If this is the case, then one should reconsider the issue of the physical origin of the three-peak structure of the $Q/I$ profile observed by \cite{trujillo_2_stenflo-keller97} in the D$_2$ line of Ba {\sc ii}, whose central peak is dominated by the $82\%$ of the barium isotopes devoid of HFS. Fig. 7 shows an interesting comparison of the magnetic sensitivity of the $Q/I$ profile of this spectral line, assuming an optically thin slab illuminated from below by the solar continuum radiation.

\begin{figure*}
\plottwo{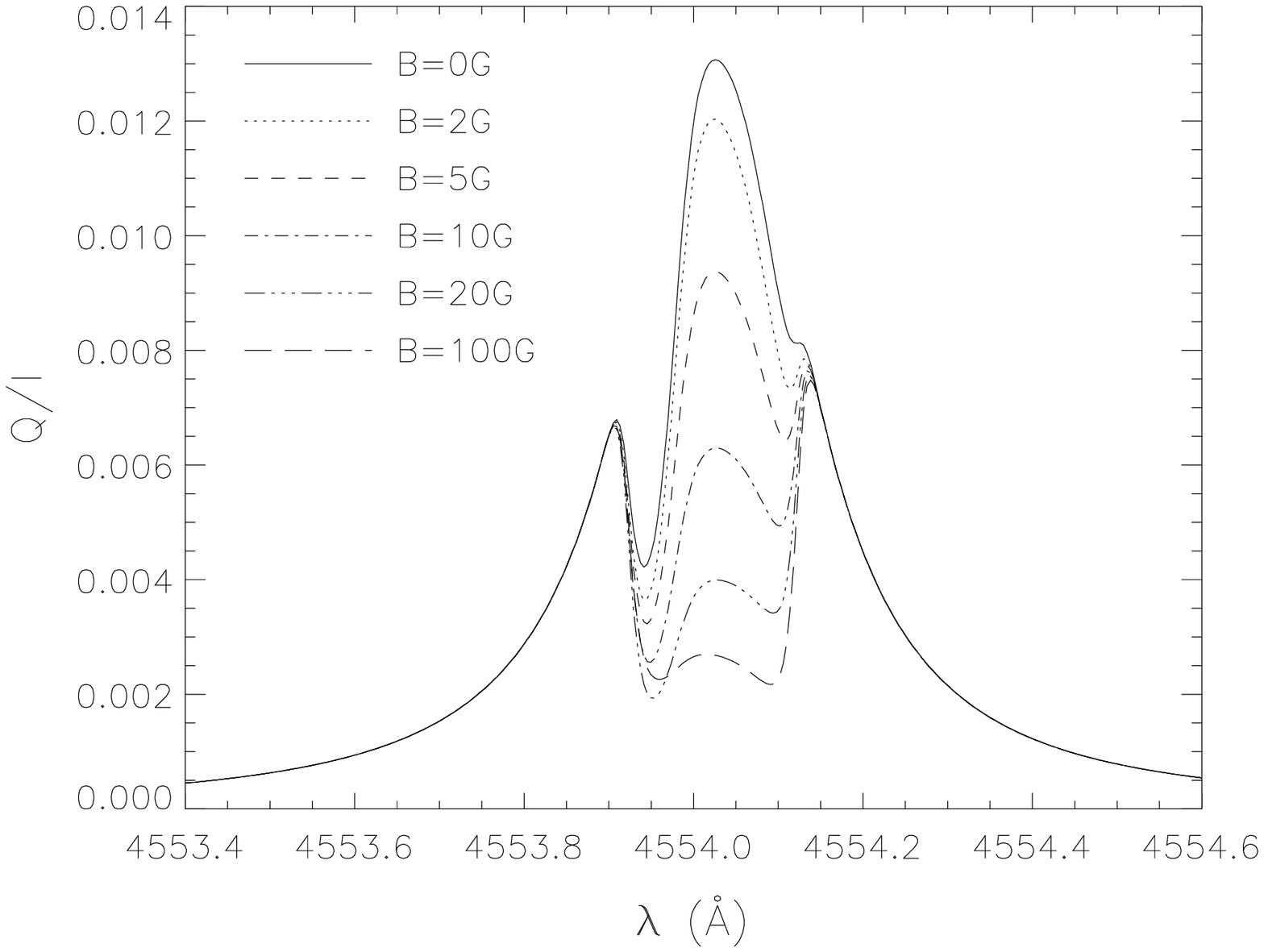}{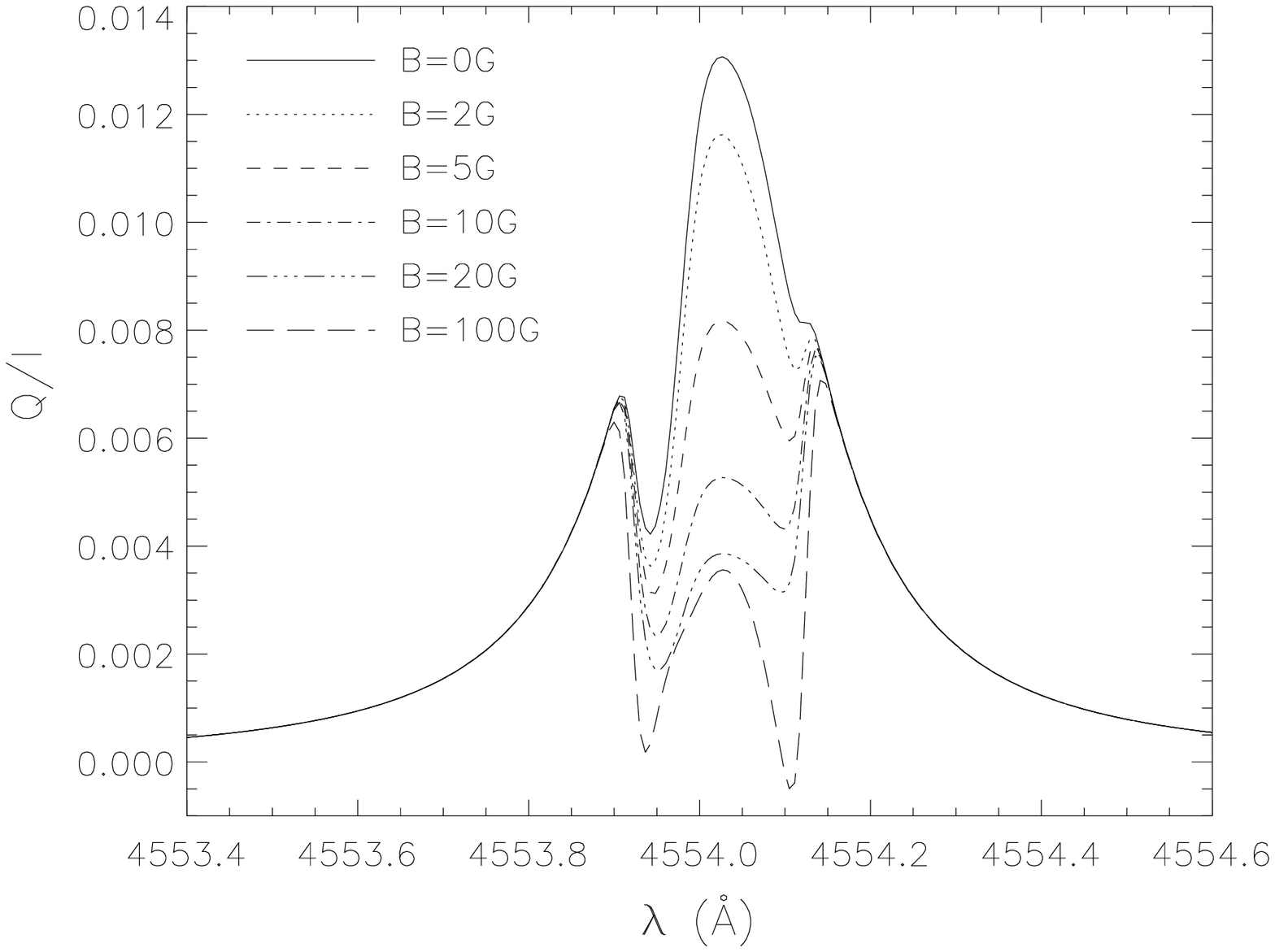}
\caption{Theoretical $Q/I$ profiles of the Ba {\sc ii} D$_2$ line in $90^{\circ}$ scattering geometry, assuming the presence of a microturbulent and isotropic field (left panel) and a horizontal field of random azimuth (right panel). See \cite{trujillo_2_belluzzis07} for a detailed, basic investigation of the polarization caused by the joint action of the Hanle and Zeeman effects in the D-lines of Ba {\sc ii}.
\label{trujillo_2_fig:barium}}
\end{figure*} 

{\bf (B) The D$_1$ Line} 

First of all, I must say that I do not see any reason to consider that the antisymmetric $Q/I$ profile observed by Trujillo Bueno et al. (2001) around the very line center of the D$_1$ line (see Fig. 2c) is an artifact produced by the polarimeter we used and/or by our data reduction strategy. Likewise, I do not have any reason to believe that the symmetric $Q/I$ profile observed by \cite{trujillo_2_stenflo-keller97} is non-solar (see the solid line of Fig. 2a). Actually, I think that both $Q/I$ features are produced by the Sun as a result of the complex dynamic and inhomogeneous nature of the ``quiet" solar atmosphere. Moreover, I even think that it should be possible to detect both $Q/I$ signals ``coexisting" in a single observation (e.g., at different positions along the spatial direction of the spectrograph's slit). But, if this conjecture turns out to be correct, which are the physical mechanisms that produce them?

The physical origin of the antisymmetric $Q/I$ profile observed by Trujillo Bueno et al. (2001) around the very line center of the D$_1$ line (see Fig. 2c) seems to be now clear. It is caused by the atomic polarization of the upper and lower HFS levels of D$_1$, with ``zero-field" dichroism playing a significant role. We saw in Fig. 3 that the anisotropy of the D-line radiation is sufficiently high so as to be able to explain the $Q/I$ amplitude we observed in the D$_1$ line. As deduced from Fig. 1 of \cite{trujillo_2_trujillos02}, the scattering polarization in the D$_1$ line is expected to be very sensitive to the strength and orientation of sub-gauss magnetic fields, but completely negligible for $B{>}10$ G  irrespective of the magnetic field inclination. Moreover, if the ground-level rates of elastic collisions with neutral hydrogen atoms used by \cite{trujillo_2_kerkeni-bommier} are really reliable, then the spatio-temporal regions of the lower solar chromosphere where the antisymmetric $Q/I$ signal originates 
(i.e., probably the lowest temperature phases of the shock-dominated region that \citeauthor{trujillo_2_rutten07}  \citeyear{trujillo_2_rutten07} calls {\em clapotisphere})
should have a hydrogen number density significantly smaller than  
$10^{14} {\rm cm}^{-3}$)

In order to constrain the physical mechanism that is behind the $Q/I$ peak observed by \cite{trujillo_2_stenflo-keller97}, it is first important to emphasize that for magnetic strengths sensibly larger than 10 G the only expected linear polarization signal in the D$_1$ line is that caused by the transverse Zeeman effect, which for $B\,{\gtrsim}\,50$ G can produce a symmetric $Q/I$ profile \citep[]{trujillo_2_trujillos02}. Some illustrative examples can be seen in Fig. 2 of \cite{trujillo_2_trujillos02} for the case of a vertical magnetic field, while Fig. 4 of \cite{trujillo_2_casini-manso} also shows the case of inclined fields with a uniformly distributed azimuth. Both calculations assumed an optically thin slab of sodium atoms, which is the reason why the vertical magnetic field case produces a {\em negative} $Q/I$ peak at the line center of the D$_1$ line, while the case of a horizontal field with a random azimuth gives instead a {\em positive} $Q/I$ peak \citep[i.e., like the one observed by][]{trujillo_2_stenflo-keller97}. It is however necessary to point out that radiative transfer effects in semi-empirical models of the solar atmosphere reverse the sign of the emergent $Q/I$, so that the transverse Zeeman effect of a uniformly distributed vertical magnetic field is actually expected to produce a symmetric $Q/I$ profile with a positive line-center peak. However, the saturation effects of the radiative transfer process at the central wavelength of the relatively strong D$_1$ line drastically reduce the central peak of the emergent $Q/I$ profile \citep[]{trujillo_2_aar-jtb-09}. For this reason, one might perhaps be inclined to think that the transverse Zeeman effect cannot be the cause of the enigmatic $Q/I$ peak observed by \cite{trujillo_2_stenflo-keller97}. What, then, could its physical origin be? 
I think that it is simply the transverse Zeeman effect produced by predominantly horizontal magnetic fields that permeate nearly {\em optically thin filamentary structures}  
located at the very top and just above the clapotisphere. Afterall, we must remember that we are dealing here with a fibrilar magnetism-dominated medium (cf. Rutten 2007), which cannot be modeled properly by any standard semi-empirical 1D model. 

Finally, it may be of interest to point out that the atomic polarization of the Na {\sc i} levels may produce observable effects even on the  emergent circular polarization of the sodium D-lines (see Fig. 8). The responsible physical mechanism is similar to that producing net circular polarization in the He {\sc i} D$_3$ multiplet observed in solar prominences \citep[see][]{trujillo_2_landi82}, namely the alignment-to-orientation transfer mechanism (which for the Na {\sc i} case is only effective when quantum interferences between the different $F$ levels are important, as shown in Fig. 2 of Casini et al. 2002).

\begin{figure*}
\plottwo{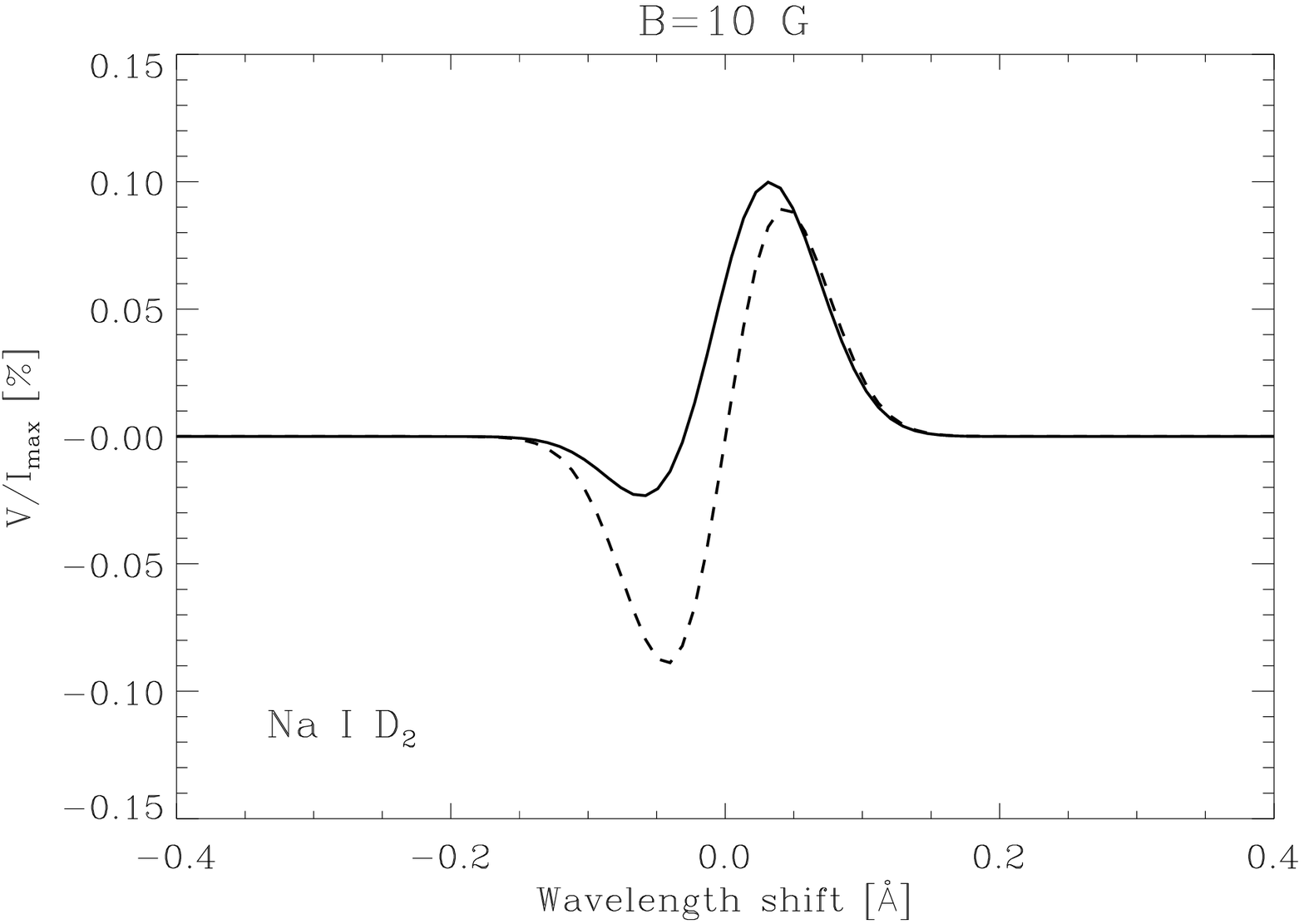}{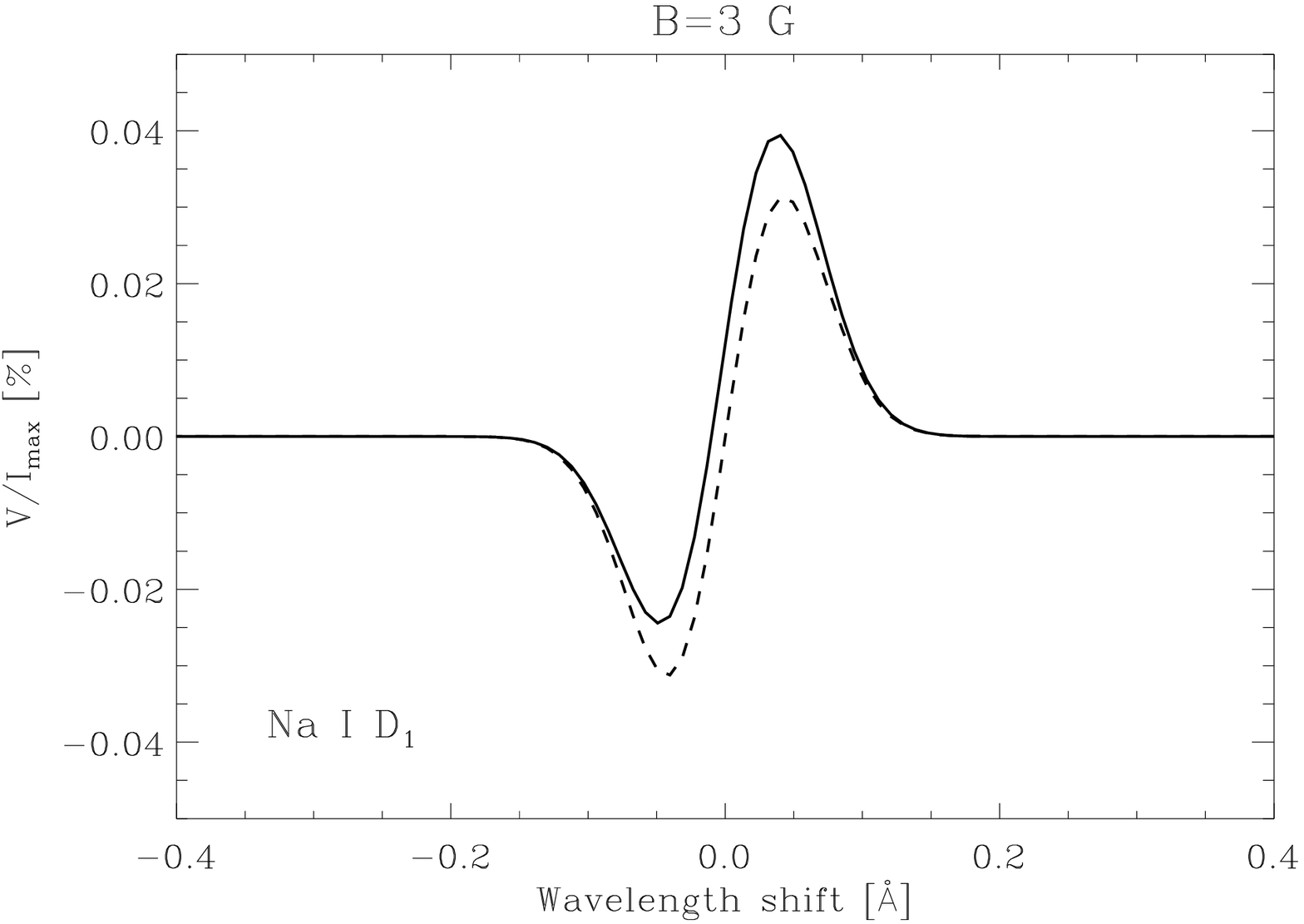}
\caption{Theoretical circular polarization profiles of the Na {\sc i} D-lines at $\mu=0.1$, assuming a thermal velocity of $3\,{\rm km}\,{\rm s}^{-1}$ and a magnetic field inclined by $\theta_B=20^{\circ}$ (with respect to the solar local vertical) and with azimuth $\chi_B=45^{\circ}$. Dashed lines: considering only the influence of the Zeeman effect. Solid lines: taking also into account the impact of the atomic level orientation that results from the atomic alignment induced by anisotropic radiative pumping processes at a height of 3" above the visible solar surface. 
\label{trujillo_2_fig:orientation}}
\end{figure*}

\section{3D Modeling of the Hanle Effect in Convective Atmospheres}

For the moment, the second solar spectrum has been observed without or with poor spatial and/or temporal resolution. For this reason, 
\cite*{trujillo_2_trujillos04} confronted observations of the center-to-limb variation of the scattering polarization in the Sr {\sc i} 4607 \AA\ line  
with calculations of the $Q/I$ profiles that result from 
{\em spatially averaging} the emergent $Q$ and $I$ profiles calculated in a three-dimensional (3D) hydrodynamical model of the solar photosphere\footnote{For the implications of a 3D radiative transfer modeling of the scattering polarization observed in MgH lines see \cite{trujillo_2_aar-jtb05}.}. The very significant discrepancy between the calculated and the observed polarization amplitudes indicated the ubiquitous existence of a hidden, unresolved magnetic field in the quiet solar photosphere. The inferred mean strength of this hidden field turned out to be $\langle B \rangle {\sim} 100$\,G, which implies an amount of magnetic energy density that is more than sufficient to compensate the energy losses of the outer solar atmosphere \citep[]{trujillo_2_trujillos04}. This estimation was obtained by using the approximation of a {\em microturbulent} field (i.e., that the hidden field has an isotropic distribution of orientations within a photospheric volume given by ${\cal  L}^{3}$, with ${\cal  L}$ the mean-free-path of the {\em line-center} photons). Calculations based on the assumption that the unresolved magnetic field is instead horizontal also lead to the conclusion of a significant amount of hidden magnetic energy in the bulk of the solar photosphere \citep*[see \S~4 in][]{trujillo_2_trujillos06}.

What is the physical origin of this hidden magnetic field whose reality is now being supported by \cite{trujillo_2_lites08} through high-spatial-resolution observations of the Zeeman effect taken with Hinode? Is it mostly the result of dynamo action by near-surface convection, as suggested by \cite{trujillo_2_cattaneo99}? Or is it dominated by small-scale flux emergence from deeper layers and recycling by the granulation flows? The fact that the inferred magnetic energy density is a significant fraction (i.e., ${\sim}20\%$) of the kinetic energy density, and that the scattering polarization observed in the Sr {\sc i} 4607 \AA\ line does not seem to be modulated by the solar cycle, suggested that a small-scale dynamo associated with ``turbulent" motions within a given convective domain of ionized gas plays a significant role for the ``quiet" Sun magnetism \citep[]{trujillo_2_trujillos04}. Recent radiative MHD simulations of dynamo action by near-surface convection also support this possibility \citep[]{trujillo_2_vogler07}.

\begin{figure}
\begin{center}
\includegraphics[width=14 cm]{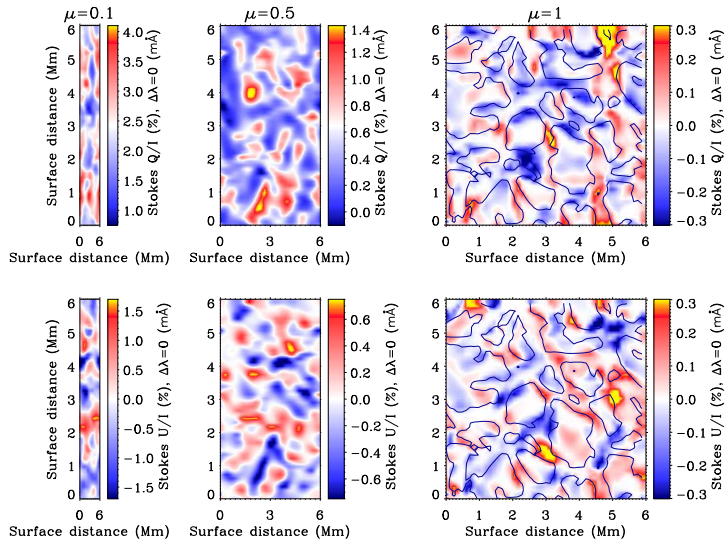}
\end{center}
\caption{The emergent $Q/I$ (top panels) and $U/I$ (bottom panels) at the line-center of the Sr {\sc i} 4607 \AA\ line calculated for three lines of sight in a 3D snapshot of a realistic hydrodynamical simulation of solar surface convection and accounting for the diffraction limit effect of a 1-m telescope. The contours in the right panels delineate the upflowing regions. From \cite{trujillo_2_trujillo-shchukin07}.}
\label{trujillo_2_fig:sr}
\end{figure}

The next step in our research on the Hanle effect in convective atmospheres will be to use snapshots from such MHD simulations in order to determine whether they can explain the ``observed" magnetic  depolarization. As shown by \cite{trujillo_2_manso-bueno99}, this problem is significantly more complicated than that considered by \cite{trujillo_2_trujillos04} because one has to take into account that at each grid-point of the computational box the corresponding magnetic field vector (with its strength, inclination and azimuth) couples all the $\rho^2_Q(J_u)$ multipolar components of the atomic density matrix among them. Recently, \cite{trujillo_2_trujillo-shchukin07} have paved the way 
towards such a goal. These authors wanted to demonstrate that 
there are further scientific reasons for observing the second solar spectrum with a spatial resolution significantly better than 1 arcsec\footnote{One of the known reasons is that a joint analysis of the Hanle effect in the Sr {\sc i} 4607 \AA\ line and in the C$_2$ lines of the Swan system indicated that the strength of the hidden field fluctuates on the spatial scales of the solar granulation pattern, with relatively weak fields in the upflowing cell centers and with 
$\langle B \rangle {\gtrsim}\, 200$\,G in the downflowing plasma \citep[]{trujillo_2_trujillos04,trujillo_2_trujillos06}.}. To this end, they solved the 3D radiative transfer problem of scattering polarization in the Sr {\sc i} 4607 \AA\ line taking into account not only the anisotropy of the radiation field in the same 3D model of solar surface convection used by \cite{trujillo_2_trujillos04}, but also the {\em symmetry-breaking} effects caused by the horizontal atmospheric inhomogeneities\footnote{Interestingly enough, this local breaking of the axial symmetry of the photospheric radiation field implies that even vertical magnetic fields can produce Hanle depolarization!}. As shown in Fig. 9, the calculated $Q/I$ and $U/I$ linear polarization signals of the emergent spectral line radiation have sizable values and fluctuations, even at the very center of the solar disk where we observe the forward scattering case (see the $\mu=1$ panels). We pointed out that the predicted small-scale patterns in $Q/I$ and $U/I$ are of great diagnostic value, because they are sensitive to the thermal, dynamic and magnetic structure of the quiet solar atmosphere. While a 1-m telescope with adaptive optics should be sufficient for detecting them in the strongly polarizing Sr {\sc i} 4607 \AA\ line, the observation of this type of linear polarization patterns in most of the other lines of the Fraunhofer spectrum would require the development of a large aperture solar telescope. For the moment, it would be interesting to investigate whether the large spatial vatiations in $Q/I$ and $U/I$ observed by \cite{trujillo_2_stenflo06} in the K-line of Ca {\sc ii} can be interpreted in terms of largely resolved magnetic fields in the solar chromosphere, or if the 
symmetry-breaking effects associated to the supergranulation network play a significant role. 

\section{Concluding Comment}

Observing the second solar spectrum with high spatial resolution would allow us to discover hitherto unknown aspects of the Sun's hidden magnetism. For this reason, the design of the European Solar Telescope (EST), and of any future space telescope (e.g., SOLAR-C), should incorporate the scientific case of the spectral line polarization produced by radiatively induced quantum coherences in atomic and molecular systems. You can be sure that the magnetic fields of the extended solar atmosphere are continuously giving rise to amazing signatures in the emergent spectral line polarization, whose observation 100 years after Hale's (1908) discovery would lead to a new revolution in our empirical understanding of solar magnetism.

\acknowledgments
{This paper is the result of a review talk in a 
workshop held in honor of Prof. Jan Olof Stenflo on the occasion of his retirement from ETH Z\"urich after 25 years of research activities on solar polarization. Jan has played a vital role in the development of this fascinating research field of 21st century astrophysics, and I would like to thank him for so many interesting and inspiring contributions. Finantial support by the Spanish Ministry of Science through project AYA2007-63881 and by the European Commission through the SOLAIRE network (MTRN-CT-2006-035484) is gratefully acknowledged. }

\end{document}